\documentclass{article}

\usepackage[utf8]{inputenc}
\usepackage[english]{babel}
\usepackage{mathtools}
\usepackage{floatrow}
\usepackage{hyperref} 
\usepackage{mathrsfs}
\usepackage{bbold} 
\usepackage{xfrac}
\usepackage{graphicx}
\usepackage{tabularx} 
\usepackage{slashed}
\usepackage{placeins} 
\usepackage{cancel}

\newcommand{\eqn}[1]{Eq.(\ref{#1})}

\title{No Landau-Yang in QCD}
\author{W. Beenakker, R. Kleiss, G. Lustermans}
\date{September 2015}

\begin{document}

{\centering

\huge \textbf{No Landau-Yang in QCD}\\[0.5cm]
\large \textbf{W. Beenakker$^{a,b,}$\footnote{W.Beenakker@science.ru.nl}, R. Kleiss$^{a,}$\footnote{R.Kleiss@science.ru.nl}, G. Lustermans$^{b,}$}\footnote{G.Lustermans@uva.nl}\\[0.5cm]

$^a$Radboud University Nijmegen, Institute of Mathematics, Astrophysics and Particle Physics, Heyendaalseweg 135,\\
NL-6525 AJ Nijmegen, The Netherlands\\[\baselineskip]

$^b$Institute of Physics, University of Amsterdam, Science Park 904 NL-1018 XE Amsterdam, The Netherlands\\[2cm]

\textbf{Abstract}\\[.5cm]

The Landau-Yang theorem, forbidding transition amplitudes between a massive spin-1 particle and two photons, is widely assumed to apply to other massless spin-1 final state particles as well. We show that this is not true in Standard Model QCD, so that for instance antisymmetric colour-octet spin-1 quarkonia can be formed by two on-shell gluons at $\mathcal{O}(\alpha_s^2)$ in perturbative QCD.
}

\newpage

\section{Introduction}
The Landau-Yang theorem \cite{landau1948,yang1949} is an old and well-established result in QED, forbidding the decay of a massive spin-1 particle into two photons. A fine example is provided by orthopositronium: since it cannot decay into two photons but only into at least three, it has a much longer lifetime than spin-0 parapositronium. Over time the Landau-Yang theorem has come to be interpreted more broadly to imply that the decay of a spin-1 particle into any two massless spin-1 particles is also forbidden, and this is used in studies of quarkonium production, such as for instance in \cite{quarkoniumarticle} under the name of the generalized Landau-Yang selection rule. In this article we present a very simple derivation of the Landau-Yang theorem based on Bose statistics, point out a loophole presented to non-Abelian gauge theories and prove that the generalized Landau-Yang theorem does not hold for at least some QCD processes.\\
We give explicit results for next-to-leading order (NLO) processes and explain the reason behind the fact that a generalized Landau-Yang theorem seems to exist at leading order (LO), but is violated at NLO, as first noted in \cite{ma2014}. Our results are general and we show that they reduce to the approximation given in the aforementioned paper when the appropriate limits are taken.

\section{The Landau-Yang theorem in QED}
Before turning our attention to the validity of an extension of the Landau-Yang theorem to the case of QCD, we first revisit the original theorem due to Landau and Yang within the framework of QED.\\
The method Yang used to obtain his result is based on the transformation properties of the annihilation and creation operators of the electromagnetic field under rotations and parity transformations. These lead to a set of selection rules which rule out the spin-1 decay into two photons. We present here a different method using the language of transition amplitudes.\\
Consider a spin-1 particle of mass $M$, momentum $P$ and polarization $\epsilon_0$ decaying into two photons with momenta $q_{1}$ and $q_{2}$ and polarizations $\epsilon_{1}$ and $\epsilon_{2}$ respectively.
This means that we have $P\cdot\epsilon_0 = q_{1}\cdot\epsilon_{1} = q_{2}\cdot\epsilon_{2} = 0$. We also define 
\begin{align}
r \equiv \frac{1}{2}(q_1-q_2),
\end{align}
so that $P\cdot r = 0$. Hence $P$ and $r$ are independent, orthogonal momenta.\\
Owing to the Ward identity we are allowed to add to any polarization vector a term proportional to its corresponding momentum. To be specific, we re-gauge $\epsilon_{1,2}$ as follows:
\begin{align}
\epsilon_i^{\mu} \rightarrow \eta_i^{\mu} \equiv \epsilon_i^{\mu} - \frac{(\epsilon_i\cdot q_k)}{(q_i\cdot q_k)}q_i^{\mu} \qquad \text{for} \qquad (i,k) = (1,2) \text{ or } (2,1).
\end{align}
These have the additional property $P\cdot\eta_{1,2} = r\cdot\eta_{1,2} = 0$. By simple exhaustion it is easy to see that the decay amplitude $\mathscr{M}$ must necessarily be of the form
\begin{align}
\mathscr{M} = a_1(r\cdot\epsilon_0)(\eta_1\cdot\eta_2) + a_2\, \varepsilon(P,\epsilon_0,\eta_1,\eta_2) + a_3(r\cdot\epsilon_0)\varepsilon(P,r,\eta_1,\eta_2),\label{QEDamplitude}
\end{align}
where $\varepsilon(p,q,k,l)$ represents the contraction of its arguments with the Levi-Civita symbol: $\varepsilon(p,q,k,l) = \varepsilon^{\mu\nu\rho\sigma}p_{\mu}q_{\nu}k_{\rho}l_{\sigma}$. The coefficients $a_i$ can apart from constants only depend on $M$.\\
The Landau-Yang result is now easily obtained by interchanging the two outgoing photons $(\eta_1 \leftrightarrow \eta_2, r\rightarrow -r)$. Under this transformation each term changes sign, while their coefficients $a_i$ do not. To obey Bose symmetry, the coefficients $a_i$ all have to be equal to zero, which is the Landau-Yang result.\\
The loophole in this argument for the case of QCD is of course that the coefficients $a_i$ contain a colour structure as well. The colour-antisymmetric part of these coefficients therefore leads to an amplitude that is Bose-symmetric and hence not forbidden. In many cases, such as the decay $Z\rightarrow gg$ involving quark triangle diagrams, the amplitude is symmetric in the gluon colours and the generalized Landau-Yang theorem holds. This has been known for a considerable time, see for instance \cite{eichten1984} where theories beyond the Standard Model are considered in which colour-antisymmetric effective interactions can be constructed. The aim of this article however is to show that also in the minimal Standard Model the generalized Landau-Yang theorem can be evaded in QCD and to give an explanation of how this works (both at LO and at NLO).

\section{The Landau-Yang theorem in QCD at LO}
We now turn to the case of QCD by replacing the photons in the previous section by gluons. As stated before, the coefficients $a_i$ now contain colour structures which might provide a loophole to evade the Landau-Yang theorem in QCD. We consider the process
\begin{align}
g(q_1,\eta_1,j)\ g(q_2,\eta_2,k) \rightarrow q(p_1,a)\ \bar{q}(p_2,b),
\end{align}
where two incoming gluons with momenta $q_{1,2}$, polarization vectors $\eta_{1,2}$ and colours $j,k$ produce a quark-antiquark pair with momenta $p_{1,2}$ and colours $a,b$. The outgoing quarks have mass $m$. At the kinematic threshold this process is equivalent to the decay of quarkonium into two gluons considered in the rest-frame.
\begin{figure}[]
\includegraphics[width=0.9\textwidth]{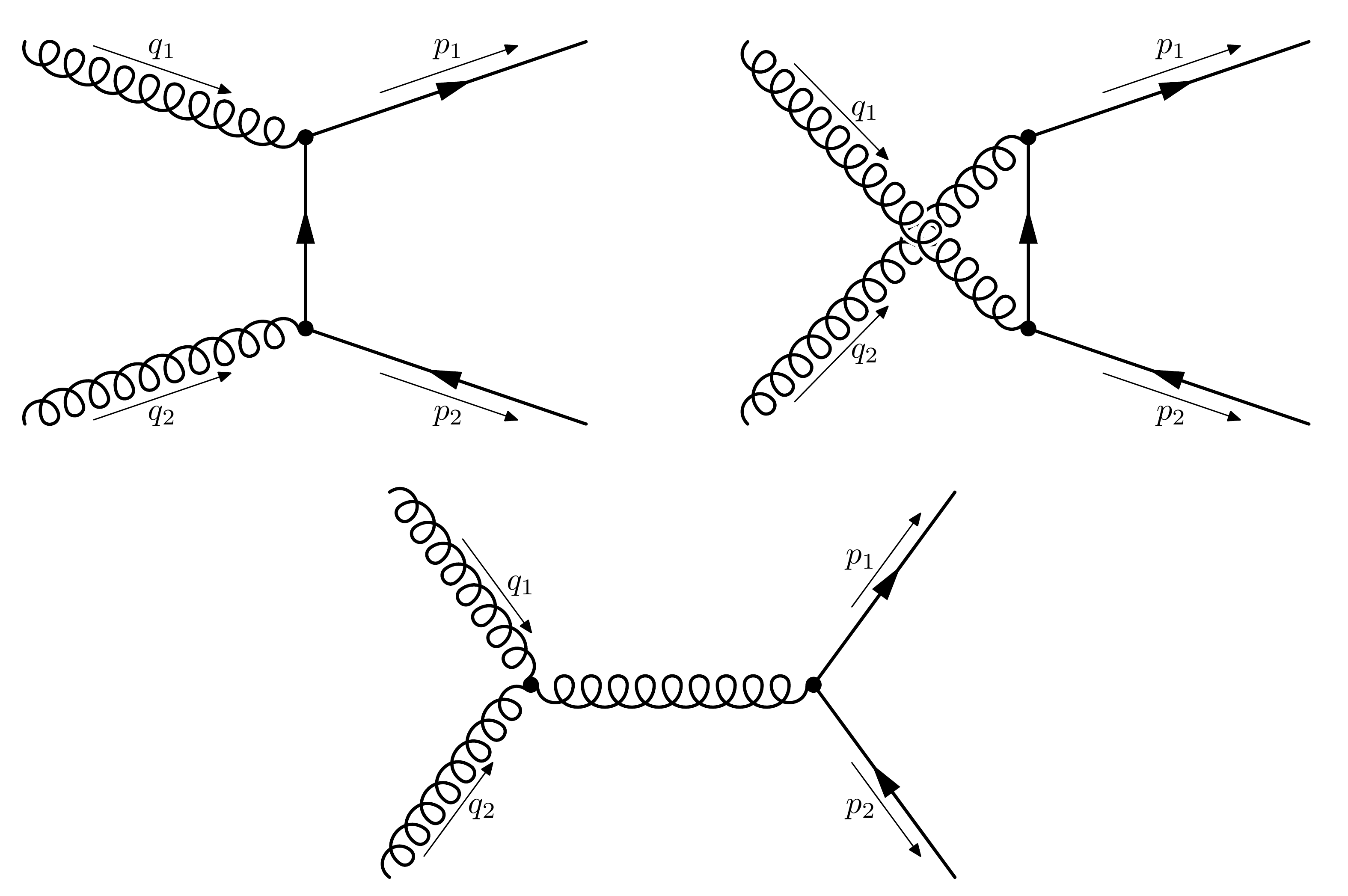}
{\caption{The diagrams contributing to $gg\rightarrow q\bar{q}$ at LO.\label{ggqqLO}}}
\end{figure}%

At LO in the strong coupling $g_s$, this involves three diagrams as depicted in \autoref{ggqqLO}. In order to discuss the amplitude, we introduce
\begin{align}
p\equiv\frac{1}{2}(q_1+q_2)=\frac{1}{2}(p_1+p_2), \qquad q\equiv\frac{1}{2}(p_1-p_2), \qquad r\equiv\frac{1}{2}(q_1-q_2).
\end{align}
By also defining the slightly altered Mandelstam variables
\begin{align}
s\equiv2(q_1\cdot q_2)=4p^2, \qquad t\equiv -2(q_1\cdot p_1), \qquad u\equiv -2(q_1\cdot p_2),
\end{align}
we can express the three diagrams referred to as the $s$-, $t$- and $u$-channel graphs, as
\begin{align}
\mathscr{M}_{0,s} &= -\frac{2ig_s^2}{s}\bar{u}(p_1)(\eta_1\cdot\eta_2)\slashed{r}v(p_2)\times C,\\
\mathscr{M}_{0,t} &= -\frac{ig_s^2}{2t}\bar{u}(p_1)\Big((q\cdot\eta_1)\slashed{\eta}_2 + (q\cdot\eta_2)\slashed{\eta}_1 - \slashed{\eta}_1\slashed{r}\slashed{\eta}_2\Big)v(p_2)\times (A+C),\\
\mathscr{M}_{0,u} &= -\frac{ig_s^2}{2u}\bar{u}(p_1)\Big((q\cdot\eta_2)\slashed{\eta}_1 + (q\cdot\eta_1)\slashed{\eta}_2 + \slashed{\eta}_2\slashed{r}\slashed{\eta}_1\Big)v(p_2)\times (A-C),
\end{align}
where $C\equiv [T^j,T^k]^a_{\ b}$ and $A\equiv \{T^j,T^k\}^a_{\ b}$ denote the commutator and anti-commutator of the colour generators $T^j$ and $T^k$ in the fundamental representation. Adding them yields the total amplitude at lowest order
\begin{align}
\mathscr{M}_0 &= \frac{ig_s^2}{2}\frac{s}{ut}\bar{u}(p_1)\ \Gamma\ v(p_2)\times\Big(A+\frac{t-u}{s}C\Big), \nonumber\label{ggqqLOamplitude}\\
\Gamma &= (q\cdot\eta_1)\slashed{\eta}_2 + (q\cdot\eta_2)\slashed{\eta}_1 - 4i\frac{m}{s}\varepsilon(p,\eta_1,r,\eta_2)\gamma^5 + \frac{t-u}{s}(\eta_1\cdot\eta_2)\slashed{r}.
\end{align}
Note that the part proportional to the commutator of the colour matrices vanishes at the kinematic threshold (where $t\rightarrow u$ and $q\rightarrow 0$) because of its prefactor $(t-u)/s$. This means that the amplitude becomes colour symmetric and therefore completely analogous to the case of QED.\\
The only surviving term at the kinematic threshold is the one involving $\gamma^5$. This structure, however, vanishes when we consider the outgoing quarks to be in the spin-1 ortho-quarkonium state, which leads to an extension of the Landau-Yang theorem to QCD at lowest order.\\

The specific factorization structure exhibited by \eqn{ggqqLOamplitude}, where the symmetric and antisymmetric colour part differ by a factor $(t-u)/s$, arises due to a cancellation between the $s$-channel and the $t$- and $u$-channels. It is this factorization structure that is responsible for the vanishing of the colour antisymmetric part of the amplitude.\\

Before performing a NLO calculation for the process in which a quark-antiquark pair is produced by two gluons, we list all possible spinor structures and categorize them according to their parity, their survival in case the two quarks form a spin-1 ortho-quarkonium state and whether they survive at the kinematic threshold or not.

\begin{table}[h]
\centering
\begin{tabular}{c|c|c|c}
Structure & Parity & Threshold & Spin-1 \\
\hline\hline
$\bar{u}(p_1) v(p_2)\ (\eta_1\cdot\eta_2)$ & + & - &  \\
$\bar{u}(p_1) v(p_2)\ (\eta_1\cdot q)(\eta_2\cdot q)$ & + & - &  \\
$\bar{u}(p_1) \gamma^5 v(p_2)\ (\eta_1\cdot\eta_2)$ & - & + &  \\
$\bar{u}(p_1) \gamma^5 v(p_2)\ (\eta_1\cdot q)(\eta_2\cdot q)$ & - & - &  \\
$\bar{u}(p_1) \slashed{\eta}_1 v(p_2)\ (\eta_2\cdot q)$ & + & - &  \\
$\bar{u}(p_1) \slashed{\eta}_2 v(p_2)\ (\eta_1\cdot q)$ & + & - &  \\
$\bar{u}(p_1) \slashed{r} v(p_2)\ (\eta_1\cdot\eta_2)$ & + & + & + \\
$\bar{u}(p_1) \slashed{r} v(p_2)\ (\eta_1\cdot q)(\eta_2\cdot q)$ & + & - &  \\
$\bar{u}(p_1) \gamma^5\slashed{\eta}_1 v(p_2)\ (\eta_2\cdot q)$ & - & - &  \\
$\bar{u}(p_1) \gamma^5\slashed{\eta}_2 v(p_2)\ (\eta_1\cdot q)$ & - & - &  \\
$\bar{u}(p_1) \gamma^5\slashed{r} v(p_2)\ (\eta_1\cdot\eta_2)$ & - & + &  \\
$\bar{u}(p_1) \gamma^5\slashed{r} v(p_2)\ (\eta_1\cdot q)(\eta_2\cdot q)$ & - & - &  \\
$\bar{u}(p_1) \slashed{\eta}_1\slashed{r} v(p_2)\ (\eta_2\cdot q)$ & + & - &  \\
$\bar{u}(p_1) \slashed{\eta}_2\slashed{r} v(p_2)\ (\eta_1\cdot q)$ & + & - &  \\
$\bar{u}(p_1) \gamma^5\slashed{\eta}_1\slashed{r} v(p_2)\ (\eta_2\cdot q)$ & - & - &  \\
$\bar{u}(p_1) \gamma^5\slashed{\eta}_2\slashed{r} v(p_2)\ (\eta_1\cdot q)$ & - & - &  \\
$\bar{u}(p_1) \big(\slashed{\eta}_1\slashed{\eta}_2 - \slashed{\eta}_2\slashed{\eta}_1\big) v(p_2)$ & + & - &  \\
$\bar{u}(p_1) \gamma^5\big(\slashed{\eta}_1\slashed{\eta}_2 - \slashed{\eta}_2\slashed{\eta}_1\big) v(p_2)$ & - & + &  \\
$\bar{u}(p_1) \big(\slashed{\eta}_1\slashed{r}\slashed{\eta}_2 - \slashed{\eta}_2\slashed{r}\slashed{\eta}_1\big) v(p_2)$ & + & + & - \\
\end{tabular}
\caption{Categorization of possible spinor structures in $gg\rightarrow q\bar{q}$}
\label{Spinorcategorization}
\end{table}

Every spinor structure can be reduced to the ones found in \autoref{Spinorcategorization} by anti-commuting the various vectors or using identities like
\begin{align}
\slashed{r}\slashed{\eta}_1\slashed{\eta}_2 &= (r\cdot\eta_1)\slashed{\eta}_2 + (\eta_1\cdot\eta_2)\slashed{r} - (r\cdot\eta_2)\slashed{\eta}_1 + \frac{i}{E^2}\slashed{p}\epsilon(p,r,\eta_1,\eta_2)\gamma_5,
\end{align}
where $E$ stands for the center-of-mass energy of any one of the particles. We have used the Pauli-identity and the fact that $p$ is the only parameter with a time-component and no space-components.\\
A plus sign in \autoref{Spinorcategorization} means that the spinor structure exhibits the desired behaviour. For parity, a plus sign means that the structure is even under a parity transformation. For the threshold entry a plus sign indicates a structure that does not vanish at the kinematic threshold. A plus sign in the final entry indicates that the specific structure does not vanish when the two quarks are considered to form a spin-1 ortho-quarkonium state.\\ 
From \autoref{Spinorcategorization} it appears that only one spinor structure is possible for our process of interest if the quarks are to form an ortho-quarkonium state at the kinematic threshold. This structure, $\bar{u}(p_1)\slashed{r}v(p_2)(\eta_1\cdot\eta_2)$, is also present in the LO amplitude but vanishes due to a prefactor $(t-u)/s$.

\section{Gluon-initiated quark-antiquark pair production amplitude at NLO}
The last term in the Born amplitude (\eqn{ggqqLOamplitude}), proportional to $(\eta_1\cdot\eta_2)\slashed{r}$, only vanishes at the kinematic threshold due to its prefactors, while the other terms vanish by themselves. This means that there might be a non-vanishing transition amplitude involving this term at higher orders.\\

To monitor divergences at the kinematic threshold, we introduce the threshold parameter
\begin{align}
\beta \equiv \sqrt{1-\frac{4m^2}{s+i\delta}}, \qquad \text{with} \qquad \delta \downarrow 0,
\end{align}
which approaches zero at the kinematic threshold. All spinor structures are expanded in terms of this parameter, as are the Mandelstam variables and other prefactors.\\


\begin{figure}[]
\includegraphics[width=0.9\textwidth]{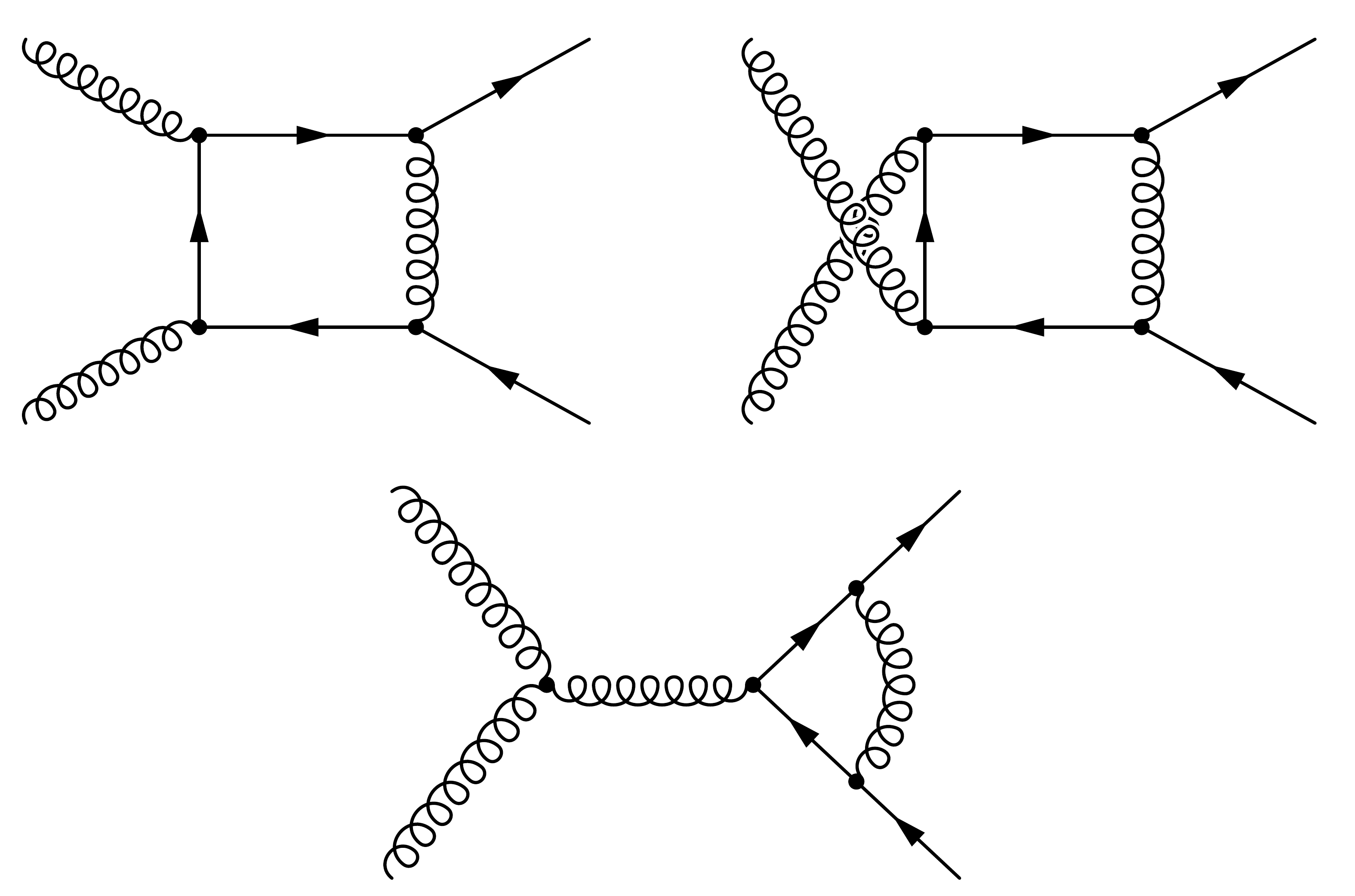}
{\caption{The Coulomb diagrams for $gg\rightarrow q\bar{q}$.\label{Coulomb}}}
\end{figure}%

To obtain the NLO amplitude for quark-antiquark pair production by two gluons, all graphs are considered at the kinematic threshold, except for the Coulomb graphs given in \autoref{Coulomb}. Diagrams containing infrared divergences, in which a soft gluon is radiated by one of the external particles, do not need to be taken into account because they can be shown to be proportional to the (vanishing) LO amplitude. The expression for each graph is then reduced to scalar integrals by the method of Passarino-Veltman reduction \cite{pvreduction,dennerpvreduction1,dennerpvreduction2}. The Coulomb graphs give rise to scalar Coulomb integrals which are proportional to $1/\beta$ and need to be expanded in terms of the threshold parameter. After all $1/\beta$ divergences are cancelled by terms arising from the expansions of the prefactors (containing $q$, $s$, $t$ and $u$) of the Coulomb integrals, the kinematic threshold limit can be taken for the Coulomb integrals as well. When all contributions are added, the fact that we are considering the quark-antiquark pair to be in the ortho-quarkonium state is taken into account and we obtain the final result\footnote{The complete calculation can be found in \cite{MasterthesisGillian}.}
\begin{align}
&\mathscr{M}_{\text{{\rmfamily\textsc{nlo}}}} = + \frac{g_s^4F}{192m\pi}\bigg[-2\ln(2\beta) + \frac{1}{\epsilon} + i\pi\bigg] \times \Big(\big\{T^j,T^k\big\}^a_{\ b} - 3\delta^a_{\ b}\delta^{jk}\Big) \nonumber\\
& \qquad \times \Big(|\eta_2|\cos(\phi_{2})\bar{u}(p_1)\slashed{\eta}_1v(p_2) + |\eta_1|\cos(\phi_{1})\bar{u}(p_1)\slashed{\eta}_2v(p_2)\Big)\nonumber\\
&+ \frac{g_s^4F}{192m^2\pi}\bigg[2\ln(2\beta) - \frac{1}{\epsilon} - 2 - i\pi\bigg]\cos(\theta) \times \bar{u}(p_1)\slashed{r}v(p_2)(\eta_1\cdot\eta_2) \nonumber\\
& \qquad \times \Big(\big\{T^j,T^k\big\}^a_{\ b} - 3\delta^a_{\ b}\delta^{jk}\Big) \nonumber\\
&+ \frac{ig_s^4}{192m^2\pi^2}\bigg[\sum_{\text{Quarks}}F'\bigg(6\beta'\ln(x'_s)\frac{m_q^2}{m^2} + 12\frac{m_q^2}{m^2} + 2 + \frac{3}{2}\ln^2(x'_s)\frac{m_q^2}{m^2}\bigg) \nonumber\\
& \qquad + F\Big(14 - \pi^2 -28\ln(2)\Big)\bigg] \times \bar{u}(p_1)\slashed{r}v(p_2)(\eta_1\cdot\eta_2)\big[T^j,T^k\big]^a_{\ b}. \label{result}
\end{align}
Here we have defined $\epsilon$ by $d = 4-2\epsilon$, where $d$ is the number of space-time dimensions. Furthermore we have introduced
\begin{align}
\beta' &\equiv \sqrt{1-\frac{4m_q^2}{s+i\delta}}, \qquad x_s' \equiv \frac{\beta'-1}{\beta'+1}, \nonumber\\
F &\equiv \left(\frac{m^2e^{\gamma_E}}{4\pi\mu^2}\right)^{-\epsilon} \qquad \text{and} \qquad F' \equiv \left(\frac{m_q^2e^{\gamma_E}}{4\pi\mu^2}\right)^{-\epsilon},
\end{align}
where $m_q$ denotes the mass of a quark occurring in a quark loop and $\mu$ indicates the `t Hooft scale. The angles $\phi_{1,2}$ refer to the angles between $p_1$ and $\eta_{1,2}$ and the angle $\theta$ denotes the angle between the incoming and the outgoing particles. Strictly speaking these are not defined at the threshold; this is a reflection of the fact that a realistic treatment would involve folding the amplitude with a quarkonium wavefunction. Secondly the colour symmetric part of the amplitude does not vanish, which might seem to contradict the Landau-Yang theorem for QED. However, from the very fact that the angles $\phi_{1,2}$ and $\theta$ occur, it can be seen that these contributions come from partial waves beyond the S-wave. Therefore the Landau-Yang theorem should not be expected to hold for these terms. 
In fact, the colour symmetric structures in \eqn{result} arise as a result of Coulomb corrections proportional to $1/\beta$ that pick out $\beta$-suppressed terms. These Coulomb corrections are formally part of the $q\bar{q}$ bound state, because they appear in the limit where the slow moving, outgoing quarks exchange a virtual gluon. Because of this, only the last part of the NLO result describes the amplitude of creating a fundamental spin-1 particle, while the other terms belong to the description of the formation of a bound state from two individual quarks. As the Landau-Yang theorem explicitly deals with fundamental spin-1 particles, we focus only on the colour antisymmetric part: 
\begin{align}
\mathscr{M}_{\text{{\rmfamily\textsc{nlo}}}}^{\cancel{LY}} = &\frac{ig_s^4}{192m^2\pi^2}\bigg[\sum_{\text{Quarks}}F'\bigg(6\beta'\ln(x'_s)\frac{m_q^2}{m^2} + 12\frac{m_q^2}{m^2} + 2 + \frac{3}{2}\ln^2(x'_s)\frac{m_q^2}{m^2}\bigg) \nonumber\\
& \qquad + F\Big(14 - \pi^2 -28\ln(2)\Big)\bigg] \times \bar{u}(p_1)\slashed{r}v(p_2)(\eta_1\cdot\eta_2)\big[T^j,T^k\big]^a_{\ b},
\end{align}
in the following discussion. The $F$ and $F'$ in this expression can in principle be set to 1, since there are no $1/\epsilon$ poles present in $\mathscr{M}_{\text{{\rmfamily\textsc{nlo}}}}^{\cancel{LY}}$.

\section{On the fate of the Landau-Yang theorem in QCD}
The term in the Born amplitude that might have provided a loophole for the Landau-Yang theorem in QCD is given by
\begin{align}
\mathscr{M}_0 &= \frac{ig_s^2}{2}\frac{s}{ut}\frac{t-u}{s}\bar{u}(p_1)\slashed{r}v(p_2)(\eta_1\cdot\eta_2) \times \frac{t-u}{s}C_O.
\end{align}
As $(t-u)/s \sim \beta$, this term is doubly suppressed at the kinematic threshold. The reason for this suppression, as stated before, is the factorization due to a cancellation between the $s$-channel and the $t$- and $u$-channel graphs.\\
We now consider the total amplitude squared up to order $g_s^8$
\begin{align}
|\mathscr{M}|^2 = \mathscr{M}_0\mathscr{M}_0^{\ast} + \mathscr{M}_0\mathscr{M}_1^{\ast} + \mathscr{M}_1\mathscr{M}_0^{\ast} + \mathscr{M}_1\mathscr{M}_1^{\ast} + \mathscr{M}_0\mathscr{M}_2^{\ast} + \mathscr{M}_2\mathscr{M}_0^{\ast}\label{amplitudesquared},
\end{align}
where the subscript denotes the order.\\
The NLO amplitude has no means of cancelling the suppressing terms of the Born amplitude, which means that the first three terms in \eqn{amplitudesquared} drop out at the kinematic threshold. Concerning the fourth term, we will only focus on the colour antisymmetric part $\mathscr{M}_{\text{{\rmfamily\textsc{nlo}}}}^{\cancel{LY}}$ as discussed in the previous section. The part that involves a sum over quarks will lead to a double sum over quarks in $|\mathscr{M}_1\mathscr{M}_1^{\ast}|$. In order for $(\mathscr{M}_0\mathscr{M}_2^{\ast} + \mathscr{M}_2\mathscr{M}_0^{\ast})$ to cancel these terms, next-to-next-to-leading order (NNLO) diagrams involving two quark loops are required. Such NNLO diagrams do not give rise to integrals containing factors of $1/\beta$ which could cancel the suppressing factors of $\mathscr{M}_0$. This means that the terms proportional to $F'$ in the NLO amplitude will be present in $|\mathscr{M}|^2$ and the Landau-Yang theorem does not hold in QCD!\\

\begin{figure}[]
\includegraphics[width=0.9\textwidth]{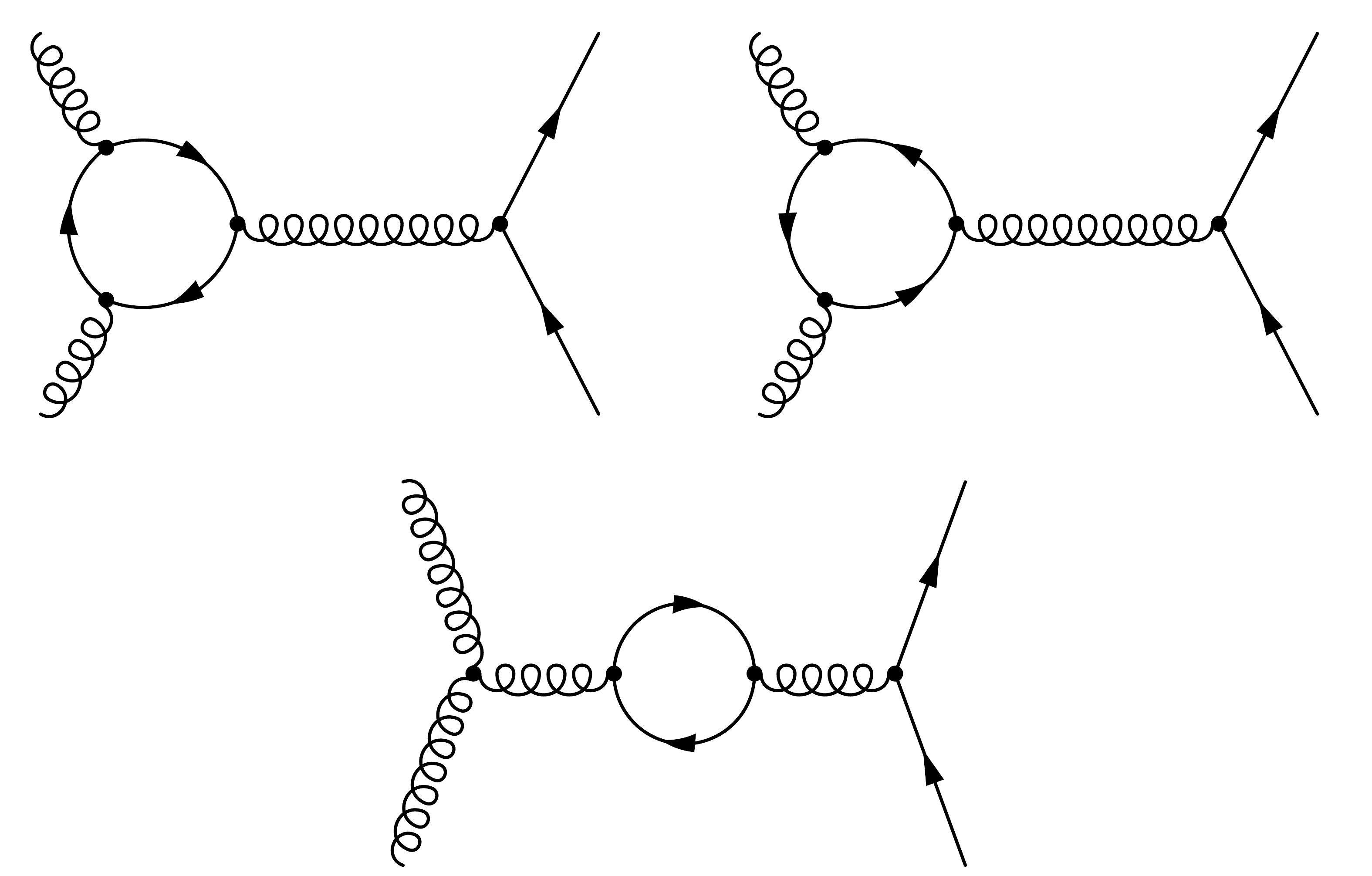}
{\caption{The NLO graphs involving quark loops.\label{quarkloops}}}
\end{figure}%

The terms proportional to $F'$ originate from the graphs involving quark loops, given in \autoref{quarkloops}. These three sets of graphs are all $s$-channel diagrams, which means that the cancellation between the $s$-channel and the $t$- and $u$-channel that gave rise to the specific factorization structure at LO does not occur for the graphs given in \autoref{quarkloops}. Thus the loophole for the Landau-Yang theorem in QCD, which was closed at LO by threshold suppression, opens up at NLO.\\

To summarize, the calculation of the 1-loop amplitude for quark-antiquark pair production by two gluons involves, among others, terms proportional to the sum over virtual quarks. We argue that, at the cross-section level, these terms can not be cancelled by terms resulting from the interference between the Born amplitude and the two-loop amplitude. As all lower order contributions vanish, we conclude that for this process, the generalized Landau-Yang theorem is violated at two-loop level in the cross-section.\\
We thus have here a mechanism to form vector quarkonia in an antisymmetric colour-octet state using on-shell gluons, which can be described by hadronic PDF's, without resorting to input from outside the Standard Model.\\

Once a particular flavour for the outgoing quark-antiquark pair is chosen, quite often an effective-field-theory (EFT) motivated approximation is employed. Heavy quarks, i.e. quarks with a mass larger than the mass of the produced quarks, are treated as decoupling. Therefore, the sum in $\mathscr{M}_{\text{{\rmfamily\textsc{nlo}}}}^{\cancel{LY}}$ only involves quark masses $m_q$ up to the mass of the produced quarks in that approximation.\\
The light quarks with a mass smaller than the mass of the produced quarks are treated as massless, causing all terms inside the sum to vanish except for the constant `$2$'. For the quark flavour equal to the flavour of the produced quarks, we have $m_q=m$, which leads to a factor of $14-3\pi^2/2$. In this EFT approximation we therefore have
\begin{align}
\mathscr{M}_{\text{{\rmfamily\textsc{nlo}},approx}}^{\cancel{LY}} = &\frac{ig_s^4}{192m^2\pi^2}\bigg[14 - \frac{3}{2}\pi^2 + 2\cdot n_l + \Big(14 - \pi^2 -28\ln(2)\Big)\bigg]\nonumber\\
& \qquad \times \bar{u}(p_1)\slashed{r}v(p_2)(\eta_1\cdot\eta_2)\big[T^j,T^k\big]^a_{\ b},
\end{align}
where $n_l$ denotes the number of light quarks.\\

The non-vanishing NLO amplitude for the process of charm-anticharm production by two gluons and the fact that the Landau-Yang theorem does not hold in this case was first noted in \cite{ma2014}. Our result (\eqn{result}) represents a more general process in which any quark-antiquark pair can be produced by two gluons and where the contributions from individual quark loops are not subject to approximations.\\
When we consider the outgoing quark-antiquark pair to be a charm-anticharm pair ($m=m_c$), our expression for the NLO amplitude indeed reduces to the result obtained in \cite{ma2014} if we treat the lighter quarks as massless and the heavier quarks as decoupling.


\end{document}